\documentstyle[graphicx,floats,prl,aps]{revtex}
\begin{document}
\draft
\twocolumn[\columnwidth\textwidth\csname@twocolumnfalse\endcsname

%\documentstyle[prc,graphicx,floats,aps]{revtex}
%\documentstyle[12pt,twoside,psfig,fleqn,espcrc1]{article}
%\begin{document}
%\draft
\title{Neutrino-induced fission of neutron-rich nuclei}

\author{E. Kolbe$^{1,2}$, K. Langanke$^{3,2}$ and G.M. Fuller $^{4,2}$ }
\address{
$^1$Departement f\"ur Physik, Universit\"at Basel, Basel, Switzerland \\
$^2$Institute for Nuclear Theory, University of Washington, Seattle, USA
\\
$^3$Institut for Fysik og Astronomi, {\AA}rhus Universitet
DK-8000 {\AA}rhus C, Denmark \\
$^4$Physics Department, University of California, San Diego, La Jolla,
CA92093-0319
}
\date{\today}
\maketitle

\begin{abstract}
We calculate neutrino-induced fission cross sections for selected nuclei
with $Z=84-92$. We show that these reactions populate the daughter
nucleus at excitation energies where shell effects are significantly
washed out, effectively reducing the fission barrier.
If the r-process occurs in the presence of a strong neutrino fluence,
and electron neutrino average energies are sufficiently high, perhaps as
a result of matter-enhanced neutrino flavor transformation, then
neutrino-induced fission could lead to significant
alteration in the r-process flow in slow outflow scenarios.
\end{abstract}

]
In this letter we calculate neutrino capture-induced
fission cross sections for heavy nuclei
associated with the r-process. Matter-enhanced
neutrino flavor transformation could enhance this process
in the supernova/compact object environments
commonly invoked as r-process sites.
Recent observations of r-process abundances in low-metallicity, old
galactical halo stars \cite{Sneden00} show patterns which agree
with the solar r-process distribution for nuclides with mass numbers $A
>   130$, but do not reproduce the solar r-process pattern for the lighter
r-process elements. In particular, these observed abundances show a peak
around mass number $A \sim 195$, which follows the solar r-process
distribution, and enhanced structures at around $A \sim 90$ and $\sim
132$ which do not follow the solar pattern.

It was recognized some time ago \cite{FM95} that $\nu_e$ capture
on heavy nuclei in the post-collapse supernova environment
would leave the daughter nuclei
at the high excitation energies characteristic of Gamow-Teller resonances.
This leaves these nuclei vulnerable to fission.
Recently, Qian has demonstrated
\cite{Qian02} that fission, induced by charged-current neutrino reactions
within this neutrino-driven wind scenario
\cite{Woosley90}, can account for the observed abundance patterns. In
this model
it is proposed that neutrino-induced fission occurs after the r-process
freezes out (i.e. all initial neutrons are exhausted) and the progenitor
nuclei decay to stability. It is further proposed \cite{Qian02} that no fission
cycling occurs during the r-process, i.e. neutrino-induced reactions are
unimportant during the r-process.

Neutrino capture-induced fission cross sections have not been calculated
before. Two aspects of nuclear physics conspire to make
this process potentially important in dense
environments with large neutrino fluxes:
(1) the weak strength distribution in the charged current (neutrino capture)
channel shows that the post-capture
daughter nucleus will be left
in a highly excited state; and (2) fission
barriers are lower at higher excitation energy.

It is expected that charged-current
reactions on r-process nuclides
will have larger partial fission cross
sections than neutral-current reactions, despite the fact that the
latter can be induced by $\nu_{\mu,\tau}$ neutrinos and their
antiparticles, which, in a core bounce supernova explosion,
might have larger average energies ($\langle E_\nu
\rangle \sim 20-25$ MeV) than the $\nu_e$ neutrinos have ($\langle E_\nu
\rangle \sim 10$ MeV). For the neutron-rich nuclei along the r-process path
neutrino capture cross
sections are quite large, as both allowed channels
(Fermi and Gamow-Teller (GT)) are governed by sum rules which scale
with the neutron excess $N-Z$ and the $\nu_e$ neutrino energy is large
enough to excite the centroids of these allowed responses.
Furthermore, the isobaric analogue state (IAS)
and the GT centroid are located at energies in the daughter nucleus
($E \sim 20-30$ MeV) which are
significantly above the fission barriers in these nuclei.
Hence, fission can represent an important,
even dominant decay mode following
neutrino-induced reactions on neutron rich nuclei. Obviously the
principal competing
decay mode is neutron emission, as neutron thresholds are also quite
low in r-process nuclei.

Our calculations of neutrino-induced reactions proceed through
two steps. First we calculate the
neutrino  cross sections as functions of excitation
energy in the final nucleus and then determine
the decay mode of the final nuclear state using a statistical approach.
The neutrino cross sections are calculated with the
random phase approximation (RPA), considering  multipoles up to $J=4$
and both parities. (See refs. \cite{Kolbe94}.)
Our RPA scheme treats proton and neutron degrees of
freedom separately and employs a partial occupancy formalism
for non-closed shell nuclei.
We adopt a zero-range Migdal force as a residual interaction.
We note that the RPA satisfies the Fermi and Ikeda sum rules,
which fix the total strength for the allowed transitions.

In the second step we calculate for each final state with well-defined
energy, angular momentum, and parity the branching ratios into the
various decay channels using the stastical model code SMOKER
\cite{Cowan91}, considering proton, neutron, $\alpha$ and $\gamma$
emission as well as fission.
The fission barriers employed here were taken from
the compilation of Howard and
M\"oller \cite{Howard80} and the neutron separation energies from the
mass table of Hilf {\it et al.} \cite{Hilf76}.
The final states in the residual nucleus
were taken from
the experimentally known levels supplemented at higher
energies by an appropriate level density formula \cite{Cowan91}.

Assuming a typical supernova $\nu_e$ neutrino spectrum, {\it i.e.},
a Ferm-Dirac spectrum with temperature $T_\nu=4$ MeV and zero
chemical potential, we have calculated the total ($\nu_e,e^-)$ cross
section and the neutrino-induced fission cross section for selected
even-even nuclei with charge numbers $Z=84-92$,
covering the range from stability all the way to
unstable r-process nuclides
%The results are shown
(Fig. 1).
%(\ref{fig:fig1}).
As the total cross sections are dominated by allowed
contributions (Fermi, GT), the cross sections increase linearly with
neutron excess within the various isotope chains, simply reflecting
the sum rules governing these two multipole transitions.
Fission is an important decay mode, in particular for the Ra, Th, and U
isotopes (Fig. 1).
%(Fig. \ref{fig:fig1}).
The
differences between the total and the fission cross sections,
are mainly accounted for by the partial $(\nu,e^- n)$ cross sections, although
for the lighter Po and Rn isotopes the decay into the gamma channel
can compete with the fission decay branch. Due to the relatively high
thresholds (Coulomb barriers),
branchings into the proton and $\alpha$ channels are
negligible. The competition between the two dominant decay modes,
neutron emission and fission, are shown in Fig. 2
%\ref{fig:fig2}
for selected
Th and U isotopes. In our calculation, fission
dominates the decay, except for the most neutron-rich nuclides shown.
We note that this calculation only considers the decay branchings in the
daughter nucleus, and does not follow multiple decays; {\it i.e.}, it
represents the \lq\lq first-chance\rq\rq\
fission cross sections \cite{Moretto72}.

\begin{figure}[ht]
    \begin{center}
      \leavevmode
      \includegraphics[width=0.95\columnwidth]{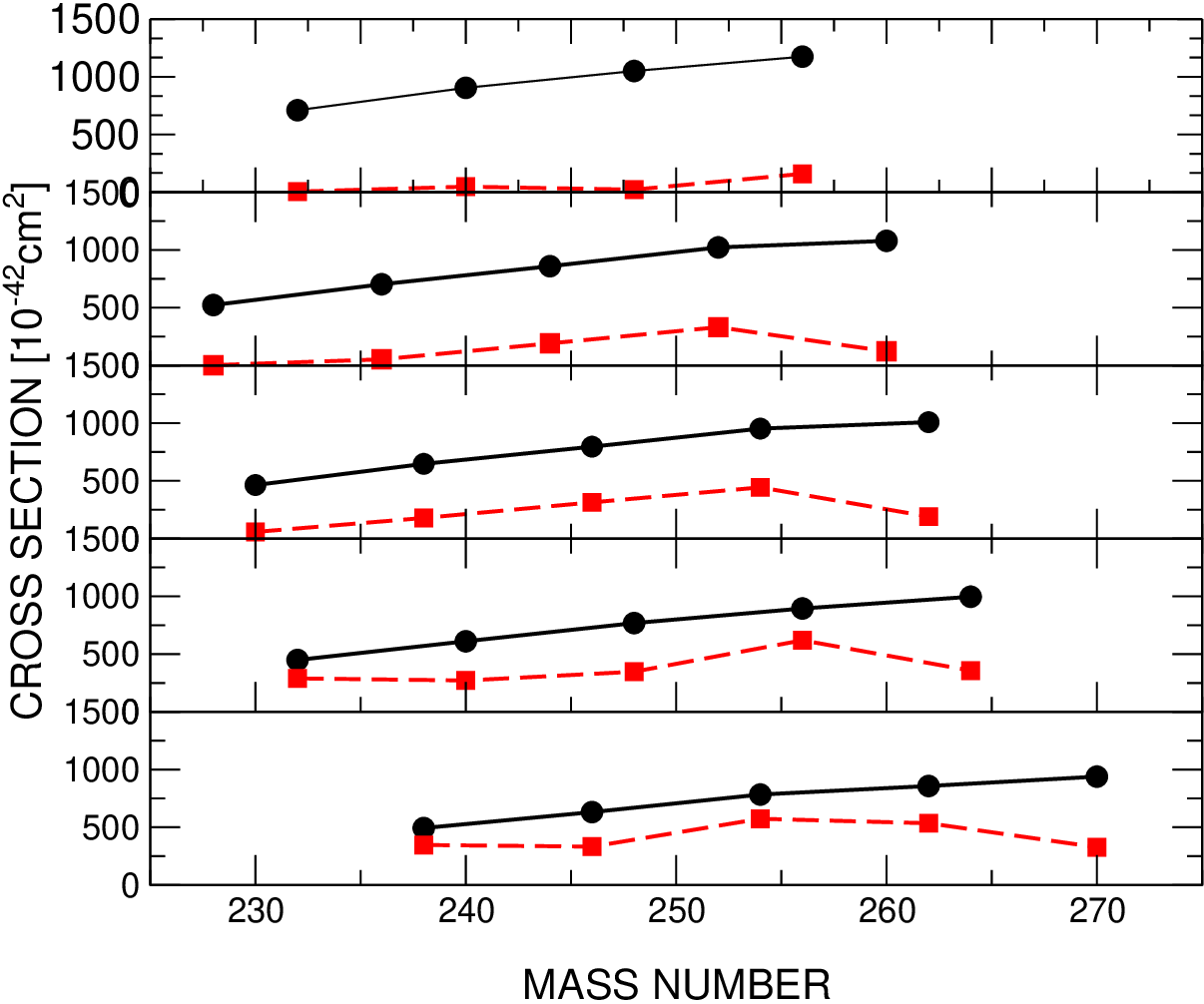}
      \caption{Neutrino-induced charged-current cross sections on selected
               Po (upper panel), Rn (second panel), Ra (third panel), Th
(fourth panel) and U (lower panel) nuclides. The total cross sections
are shown by circles and the partial fission cross section by squares.
A Fermi-Dirac
spectrum with temperature $T_\nu=4$ MeV and zero chemical potential has been
assumed for the $\nu_e$ neutrinos.
          }
      \label{fig1}
    \end{center}
\end{figure}

\begin{figure}[ht]
    \begin{center}
      \leavevmode
      \includegraphics[width=0.95\columnwidth]{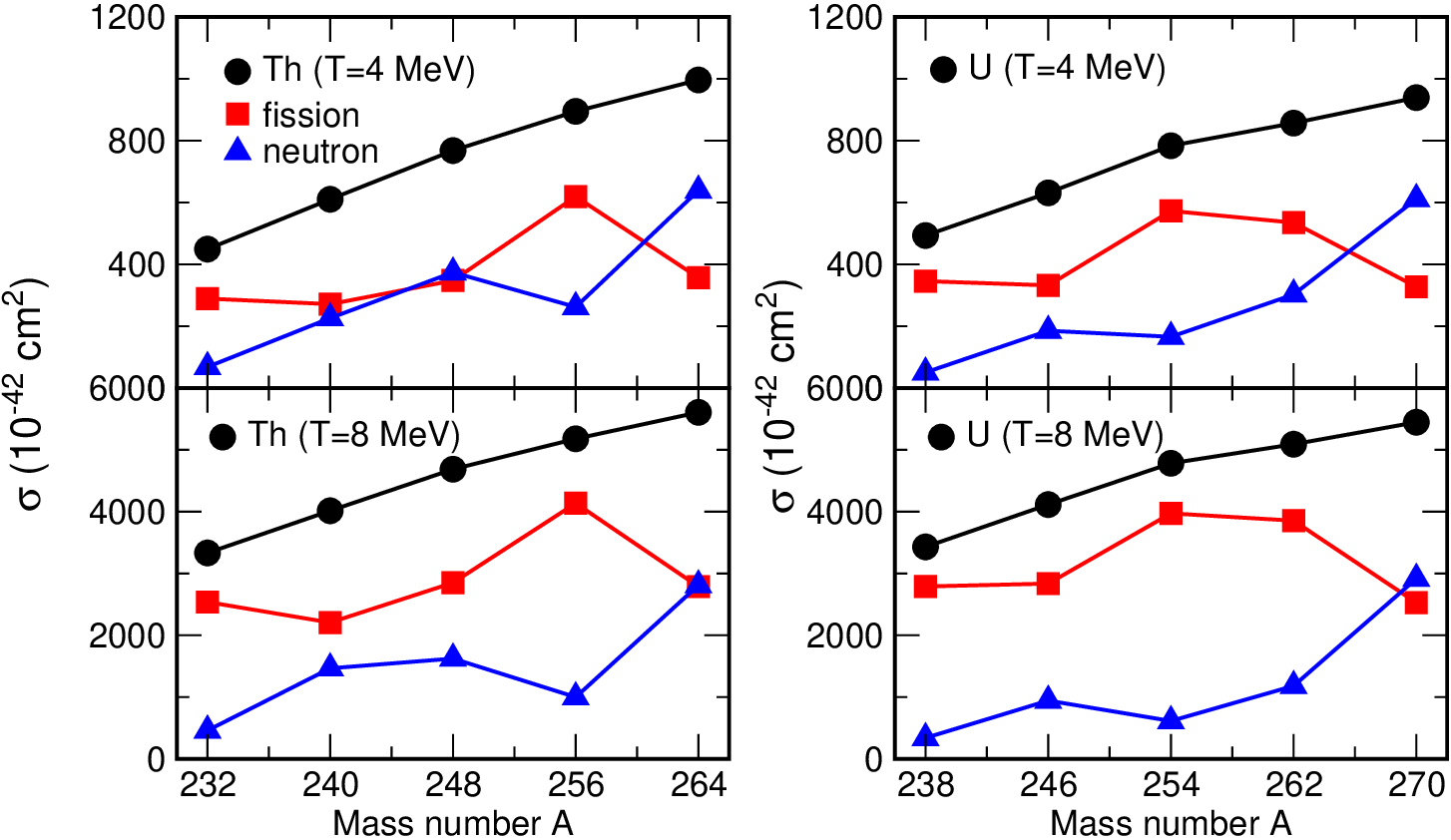}
      \caption{
Total $(\nu_e,e^-)$ (circles) and partial $(\nu_e,e^- n)$ (triangles)
and neutrino-fission cross sections (squares) for selected
Th (left panels) and U (right panels) nuclides.
The calculations have been performed for Fermi-Dirac neutrino
spectra with temperature $T_\nu=4$ MeV and 8 MeV and zero chemical potential.
          }
      \label{fig2}
    \end{center}
\end{figure}

The competition between the dominant decay modes
(fission, neutron emission) in a neutrino capture-excited daughter
is governed by the relative values of the fission barrier $B_f$ and the
neutron separation energy $S_n$. The fission probability $P_f$ is then
approximately given by \cite{Jensen79}
\begin{equation}
P_f = \frac{1}{1+ 4 (m_n/\hbar^2) R^2 T {\rm exp} \left\{(B_f-S_n)/T
\right\}}
\end{equation}
where $m_n$ is the nucleon mass and $R=1.2 \cdot A^{1/3}$ is the nuclear
radius. This formula assumes that the decaying nucleus is excited at
energies $E$ which are significantly larger than $B_f$ and $S_n$. This
is true in $(\nu_e,e^-)$ reactions with supernova neutrinos
on heavy neutron-rich nuclei where
$E \sim 25$ MeV. Such excitation energies
correspond to nuclear temperatures of $T \approx 1$ MeV in nuclei with
$A \sim 230-270$. For simplicity we have assumed that $T=1$ MeV in the
following. Eq. (1) yields
$P_f \sim 1/6$ if $B_f=S_n$, and $P_f=0.5$ if $B_f-S_n = -1.6$ MeV.
The difference $U=B_f-S_n$ is strongly dependent on
the excitation energy. In fact, Eq. (1) is derived from statistical
considerations involving the level density at vanishing nuclear
deformation (for the neutron emission probability) and at the saddle
points of the double-humped fission barriers. The latter corresponds to a
sizable nuclear deformation, where the level density increases faster
than at vanishing deformation. This reduces
$U$ with increasing excitation energy, enhancing
the fission probability relative to neutron emission
\cite{Moretto72,Jensen73}. This is consistent with the fact that
the fission barrier in heavy nuclei is
strongly influenced by shell effects \cite{Brack80,Jensen79},
and these are washed out
with increasing excitation energy.

Using Eq. (1) we have inverted our
calculated fission probabilities to obtain $U(T) = B_f (T) - S_n (T)$,
assuming $T=1$ MeV. The desired quantity $\Delta U(T) = U(T=0)-U(T)$ is
plotted in Fig. 3,
%\ref{fig:barr},
where $U(0)$ has been derived from the
compiled fission barriers \cite{Howard80} and neutron separation
energies \cite{Hilf76}. We note that the energy reduction is
significant, amounting to about 4 MeV on average. This result is in good
agreement with earlier estimates for heavy nuclei \cite{Moretto72}.
Although Fig. 3
%\ref{fig:barr}
shows some scatter among the studied
nuclei caused by structure effects, we will
assume that $U(T)$ for supernova $(\nu_e,e^-)$ reactions
on neutron-rich nuclei is lowered by 4 MeV compared to
its ground state value. This allows
for some interesting conclusions, which are rather independent of the
chosen fission barriers and neutron separation energies.

\begin{figure}[ht]
    \begin{center}
      \leavevmode
      \includegraphics[width=0.95\columnwidth]{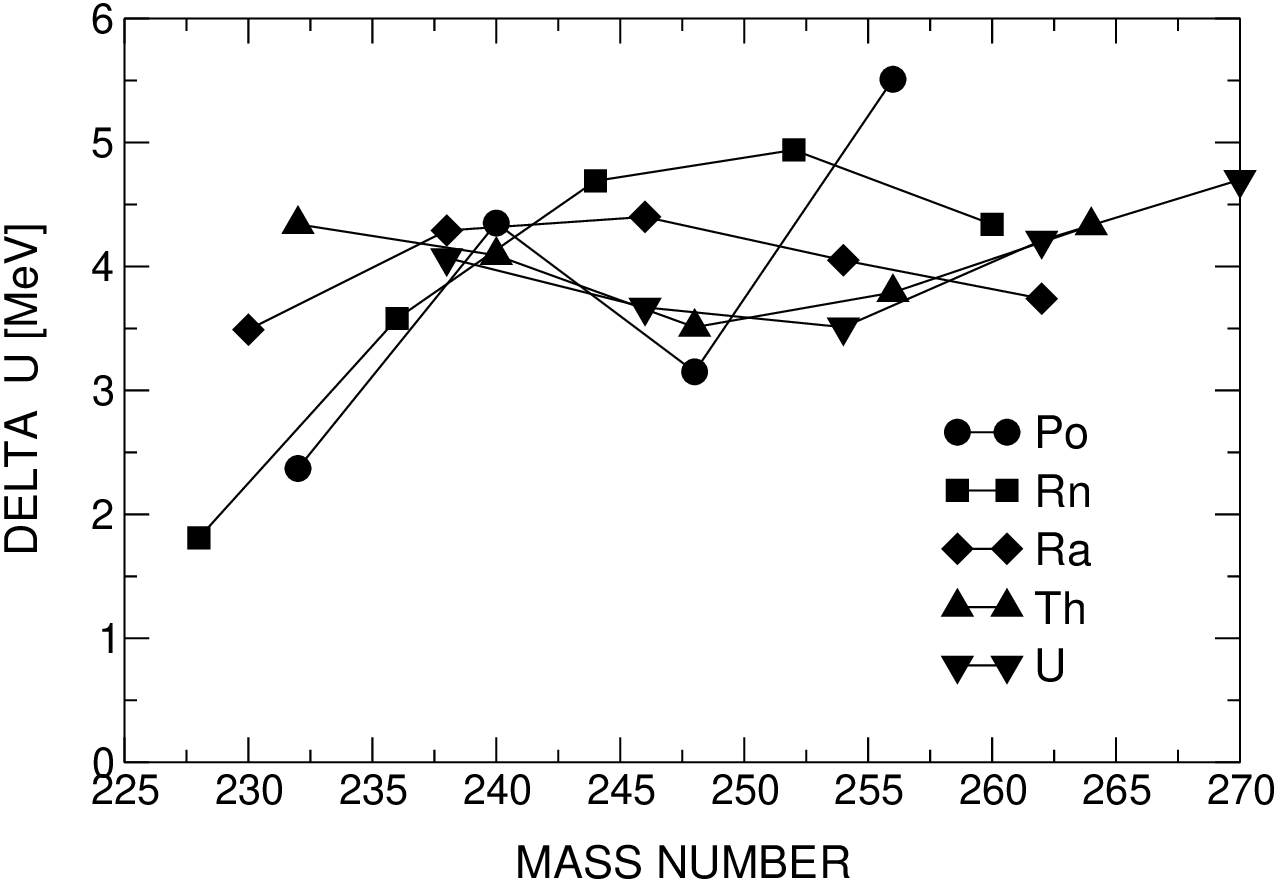}
      \caption{
The difference $\Delta U = U(0)-U(T)$
for neutrino-induced fission by $\nu_e$
neutrinos with a Fermi-Dirac distribution of temperature $T_\nu=4$ MeV and
zero chemical potential. $U(T)= B_f(T)-S_n(T)$
is calculated from our partial cross
sections using Eq. (1) and assuming a nuclear temperature $T=1$ MeV,
while $U(0)$ is derived from the tabulated fission barriers and neutron
separation energies.
          }
      \label{fig5}
    \end{center}
\end{figure}

For charged-current reactions with supernova $\nu_e$ neutrinos one then has
$P_f=0.5$ for $B_f-S_n \sim 2.4$ MeV and $P_f=0.2$ for $B_f-S_n =3.7$
MeV,
where $B_f$ and $S_n$ are the tabulated values appropriate for low
excitation energies.
Note that the predicted fission barrier heights vary
quite significantly where modern evaluations (i.e. \cite{Mamdouh00})
give higher barriers for neutron-rich nuclei than did earlier work
(i.e. \cite{Howard80}).
The recent fission barriers of
\cite{Mamdouh00} predict a fission probability $P_f > 0.2$
for most nuclei with $Z>92$ and $A>230$, with the exception of nuclei with
lower $Z$-values around the potentially magic neutron number $N=184$.
The fission barriers of M\"oller and Howard \cite{Howard80} 
allow for significant spontaneous fission probabilities for nuclei with
$Z>87$ and $A>230$, including those around $N=184$.
However,  nuclei with $A<230$ have a
smaller fission probability in neutrino-induced reactions than
tentatively assumed by Qian \cite{Qian02}.

Nuclei on the r-process path have $S_n \sim 1.5-2.5$ MeV. Such nuclei
fission after excitation by neutrinos with probability $P_f >0.2$ if $B_f \sim
5.2-6.2$ MeV. This condition is satisfied for some nuclei on the 
r-process path
with $Z \geq 96$ for the fission barriers of \cite{Mamdouh00} and for
$Z \geq 88$ for those of \cite{Howard80}, which are, however, likely too
small for neutron-rich nuclei \cite{Cowan91}. This would imply that
neutrino-induced reactions can initiate a fission cycle, if the
r-process production of superheavy elements occurs in a noticeable
neutrino fluence. Such a scenario might be conceivable, if neutrino
oscillations occur. In contrast, the fission barriers of \cite{Mamdouh00}
are too high to allow for fission cycling during the r-process
by $\beta$-delayed or neutron-induced fission.

We note that the typical fission cross section for
Th and U isotopes ($\sim 400 \cdot 10^{-42}$ cm$^2$) corresponds to a halflife
of $\sim 0.08$ s, assuming a neutrino reaction at a radius of 100 km
above the neutron star and a typical supernova $\nu_e$ luminosity of
$10^{52}$ erg s$^{-1}$. Such a halflife is shorter than the expected
halflives for the  r-process  waiting point nuclei with
$N=126, A >195$
\cite{rmp} (and also with $N=184, A > 280$ \cite{Moeller96}).
These typical halflives are also
comparable and may be shorter than the typical
$\sim 0.1\,{\rm s}$ expansion timescale in \lq\lq slow\rq\rq\
neutrino-driven wind models.
Thus, if the r-process occurs in a strong neutrino fluence
neutrino-induced fission on the progenitor nuclei during the decay to
stability
might affect the relative Th/U
r-process abundance. This abundance ratio is a necessary theoretical
ingredient if one wants to deduce an age limit for the universe from the
recently observed Th/U abundance ratios in old galactical halo stars
\cite{Cayrel01}.

The leverage that neutrino
capture-induced fission has in an r-process set in a
neutrino-driven wind is dependent on the $\nu_e$ energy
spectrum and on the neutrino fluxes
at the position where the neutrons are captured.
Models with an extremely fast outflow rate
(\cite{CF97,TB01}) generally have neutron capture occuring
far from the neutron star where neutrino fluxes are low and, hence,
neutrino capture-induced fission effects could be scant,
though post-processing fission could still be significant.

Models with a slow outflow rate suffer
from a deficit of neutrons \cite{FM95} associated with the \lq\lq 
alpha effect.\rq\rq\
However, these models can yield a viable r-process close to the neutron star
if neutrino flavor mixing effects are invoked (\cite{MFBF01,FMBF02}).
A hierarchical neutrino energy spectrum, one where the mu and tau flavor
neutrinos are more energetic than the electron neutrinos remains
a possibility for at least some epochs following the bounce of the supernova
core. In this case, matter-enhanced neutrino flavor transformation can
play an important role in determining the efficacy of neutrino
capture-induced fission, by making the average energies of the electron
neutrinos larger and, hence, boosting fission probabilities
in the region where neutrons are being captured in the r-process.
(This is demonstrated in Fig. 2 for a T=8 MeV neutrino spectrum.)
Crudely, the relationship between radial distance $r_6$ from the neutron star's
center in units of $10\,{\rm km}$ and the temperature $T_9$ in billions of
Kelvins is $r_6 \approx 22.5/(S_{100} T_9)$, where $S_{100}$ is the entropy
per baryon in units of hundreds of Boltzmann's constant. Typically,
neutron capture in the \lq\lq slow outflow\rq\rq\ schemes takes place 
in the region
where  $1 < T_9 < 3$. The location where a neutrino
of energy $E_r$ will transform from muon/tau-flavor to electron
flavor (and {\it vice versa}) is
\begin{eqnarray}
T_9^{\rm MSW} &\approx& 1.3{\left({{20\,{\rm MeV}}/{E_r}}\right)}^{1/3}
{\left({{0.42}/{(Y_e + Y_{\nu})}}\right)}^{1/3} \\ \nonumber
&\times&
S_{100}^{1/3}
{\left({{3 \times{10}^{-3}\,{\rm eV}^2}/{\delta m^2 \cos2\theta}}\right)}^{1/3}
\label{msw}
\end{eqnarray}
where $\delta m^2$ is the relevant difference of the squares of the
vacuum neutrino mass eigenvalues (scaled here by the atmospheric
neutrino value)
and $\theta$ is the effective two-neutrino vacuum mixing angle, which for
the ${\nu}_e \rightleftharpoons{\nu}_{\mu/\tau}$ transformation channel
in a strictly three-neutrino mass/mixing scheme would
be roughly ${\theta}_{1 3} < 0.15$. The experimental upper limit on this
mixing angle precludes an adiabatic transformation at resonance in a
straight MSW
scheme, but we note that large effective matter-mixing could occur
(depending on entropy and overall neutrino luminosity) on account
of the flavor basis off-diagonal
neutrino-neutrino forward scattering contributions to the weak
potential which determines neutrino effective masses and matter
mixing angles.
In this equation $Y_e$ is the electron fraction and
$Y_\nu$ is the effective neutrino number fraction which enters into
the neutrino forward scattering potential. Note that $Y_\nu$ can be negative
when the neutrino background is important.
In fact, the neutrino background potential can lead
to near maximal flavor mixing in medium. Either way, the above
expression and the
implied location of significant flavor transformation is conservative:
we may actually have a more energetic electron neutrino spectrum
on account of "chaotic" maximal mixing well below this position.
Crudely, the neutrino flux is $(5\times{10}^{42}\,{\rm cm}^{-2}\,{\rm s}^{-1}
r_6^{-2}{\left(10\,{\rm MeV}/{\langle
E_{{\nu}_e}\rangle}\right)}L_{\nu_e}^{51}$.
Here $L_{\nu_e}^{51}$ is the effective electron neutrino luminosity
in units of ${10}^{51}\,{\rm ergs}\,{\rm s}^{-1}$. If the entropy
per baryon is $S_{100}=2$, then the radius
where a neutrino of energy $E_r = 20\,{\rm
MeV}$ transforms
is $r_6 \approx 7$ (corresponding to $T_9 \approx 1.6)$ and we would
expect the typical
lifetime against fission per
big nucleus (inverse fission rate)
to be $\lambda_{\rm f}^{-1} \approx 0.05\,{\rm s}/L_{\nu_\alpha}^{51}$
where $\alpha = e,\mu,\tau$ is the flavor of the progenitor of the
electron neutrino when it leaves the neutrino sphere;
whereas, if $Y_e + Y_\nu =0.1$,
a possibility if neutrino mixing has been augmented by the neutrino background
potential(s), then $r_6 \approx 4.3$ and $T_9 \approx 2.6$ (for $S_{100} =2$)
so that $\lambda_{\rm f}^{-1} \approx 0.02\,{\rm s}/L_{\nu_\alpha}^{51}$.
In either case,
these lifetimes are shorter than typical waiting point
r-process beta decay lifetimes and are shorter than at least a plausible
range of expansion time scales, $\tau_{\rm dyn} \sim .015\,{\rm s}$.
This implies that the neutron capture flow could proceed out to
some threshold nuclide mass in the 195
peak or just beyond, whereupon fission sets in, producing two fission fragments
in the 130 peak, as outlined by Qian.

To establish a steady state fission cycling
scenario with neutrino capture-induced fission alone is problematic.
If, in steady state flow, every seed nucleus is brought by
neutron capture to a nuclear mass where the fission cross section
is greater than some threshold value, $\sigma_f^{\rm th}$, then fission
of this nucleus will result. Over a time $\Delta t \sim 2\tau_{\rm dyn}$
there will be only some
$\sim 72 \left({\lambda_{\rm f}}/{300\,{\rm s}^{-1}}\right)
\left({\Delta t}/{0.03\,{\rm s}}\right)
\left({N}/{8}\right)$ neutrons liberated
per threshold nuclear mass, where $N$ is the assumed number of neutrons
liberated per fission. Sustaining steady state fission cycle flow
would require the liberation of
some $70$ to $100$ neutrons
per fission fragment (mass $\sim 130$) and this
is clearly untenable. Nevertheless, a more modest number of
neutrons liberated per fission coupled with the
large rate of mass $130$ fission fragment production
could represent a significant alteration in the r-process flow.
At the very least it shows that the mass $130$ and $195$ peaks should
have comparable abundances.

Discussions with Aksel S. Jensen are highly appreciated. We thank F.-K.
Thielemann for making the SMOKER code available to us. This work 
is supported in part by the Danish Research Council
and by NSF Grant PHY00-99499 at UCSD.

\end{document}